\documentclass[journal,10pt]{IEEEtran}
\usepackage{enumerate}
\usepackage{cite}
\usepackage{subfigure}
\usepackage{amsmath,amssymb,amsfonts}
\usepackage{graphicx,epsfig,epstopdf,balance}
\usepackage{algorithmic}
\usepackage{graphicx}
\usepackage{xcolor}
\usepackage{textcomp}
\usepackage{float}
\usepackage{multirow}
\usepackage{stfloats}
\usepackage{chngpage}
\usepackage[ruled]{algorithm2e}
\usepackage[T1]{fontenc}

\begin{document}

\title{\fontsize{23.5}{28}\selectfont Jamming Identification with Differential Transformer \\ for Low-Altitude Wireless Networks  } 
\author{Pengyu Wang,
	Zhaocheng Wang,~\IEEEmembership{Fellow,~IEEE},
	Tianqi Mao,
	Weijie Yuan,~\IEEEmembership{Senior Member,~IEEE},\\
	Haijun Zhang,~\IEEEmembership{Fellow,~IEEE},
	and George K. Karagiannidis,~\IEEEmembership{Fellow,~IEEE}
\IEEEcompsocitemizethanks{
\IEEEcompsocthanksitem 
This work was supported in part by the National Natural Science Foundation of China under Grant U22B2057, in part by the Postdoctoral Fellowship Program of China Postdoctoral Science Foudation (CPSF) under Grant GZB20240387, in part by the China Postdoctoral Science Foundation under Grant 2024T170492. (\textit{Corresponding author: Tianqi Mao.})\par
P. Wang and Z. Wang are with the  Department of Electronic Engineering, Tsinghua University, Beijing 100084, China. Z. Wang is also with the Tsinghua Shenzhen International Graduate School, Shenzhen 518055, China
(e-mails: wangpengyu@mail.tsinghua.edu.cn; zcwang@tsinghua.edu.cn).\par
T. Mao is with the State Key Laboratory of Environment Characteristics and Effects for Near-space, Beijing Institute of Technology, Beijing 100081, China, and also with the MIIT Key Laboratory
of Complex-field Intelligent Sensing, Beijing Institute of Technology, Beijing
100081, China (e-mail: maotq@bit.edu.cn).\par 
W. Yuan is with the Department of Electrical and Electronic Engineering,
Southern University of Science and Technology, Shenzhen 518055, China
(e-mail: yuanwj@sustech.edu.cn).\par
H. Zhang is with the Beijing Engineering and Technology Research Center for Convergence Networks and Ubiquitous Services,
University of Science and Technology Beijing, Beijing 100083, China (e-mail:
zhanghaijun@ustb.edu.cn).\par
G. K. Karagiannidis is with Department of Electrical and Computer Engineering, Aristotle University of Thessaloniki, Greece (e-mail: geokarag@auth.gr).\par 

P. Wang and Z. Wang contributed equally to the project and should be
considered as co-first authors. 
}}

\markboth{}{}
\maketitle

\begin{abstract}
Wireless jamming identification, which detects and classifies electromagnetic jamming from non-cooperative devices, is crucial for emerging low-altitude wireless networks consisting of  many drone terminals that are highly susceptible to electromagnetic jamming.  However, jamming identification schemes adopting  deep learning (DL) are vulnerable to attacks involving carefully crafted adversarial samples, resulting in inevitable robustness degradation. To address this issue, we propose a differential transformer framework for wireless jamming identification.
Firstly, we introduce a differential transformer network in order to  distinguish jamming signals, which overcomes the attention noise when compared with its traditional counterpart by performing self-attention operations in a differential manner.
Secondly, we propose a randomized masking training strategy to improve network robustness, which  leverages the patch partitioning mechanism inherent to transformer architectures in order to create parallel feature extraction branches. Each branch operates on a distinct, randomly masked subset of patches, which  fundamentally constrains the propagation of adversarial perturbations across the network. Additionally, the ensemble effect generated by fusing predictions from these diverse branches demonstrates superior resilience against adversarial attacks.
Finally, we introduce a novel consistent training framework that significantly enhances adversarial robustness through dual-branch regularization. Simulation results demonstrate that our proposed methodology is superior to existing methods in boosting robustness to adversarial samples.
\end{abstract}
\begin{IEEEkeywords}
Wireless jamming identification, anti-jamming, cognitive radio, differential transformer,  low-altitude wireless network.
\end{IEEEkeywords}

\IEEEpeerreviewmaketitle

\section{Introduction}
\label{sec:introduction}

The rapid advancement of drone  technologies in recent years has prompted the rise of  low-altitude economy as a significant economic paradigm shift.  Various applications such as precision aerial imaging, automated logistics, intelligent agricultural management, and emergency response operations are outlined \cite{b1,b2}.  The evolution of low-altitude economy into a major growth driver for global economy underscores a critical reliance on advanced drone communication technologies, given its expanding operational scope and complex application scenarios.

The effective execution of low-altitude economy relies significantly on strong and high-performance wireless communication capabilities.  Stable and low-latency connectivity is essential for ensuring the operational safety of drones, facilitating real-time data exchange, and enabling reliable remote command and control.  In this context, low-altitude wireless networks (LAWNs) play a crucial role as the foundation of drone-centric communications \cite{b3}, which facilitate sustained connectivity and address the technical challenges inherent to low-altitude scenarios. 

The exponential proliferation of  drones in low-altitude wireless networks has created  unprecedented security challenges, particularly from sophisticated electromagnetic jamming threats  \cite{b4,b5}.  Drones are usually taking  mission-critical operations, including infrastructure surveillance and emergency response. Their reliance on wireless information exchange exposes them to risks from malicious jamming and unintentional interference in crowded spectral environments.  Low-altitude operations differ from traditional terrestrial networks since they are particularly vulnerable to targeted electromagnetic jamming attacks due to their line-of-sight propagation characteristics.

Electromagnetic jamming occurs when passively received signals interfere with traditional signal transmission \cite{b6}.  Specialised jamming equipment frequently produces such kind of interference, which can significantly disrupt the communication systems.  Low-altitude wireless systems have to confront various jamming sources, requiring the advancement of multiple anti-jamming technologies.  Alongside standalone anti-jamming techniques, several integrated anti-jamming technologies are usually utilized, providing the improved spectral efficiency and reliable performance \cite{b7,b8,b9}.  To guarantee the performance in the presence of complex and diverse jamming signals, specific anti-jamming strategies are essential.  It is evident that jamming identification technology is crucial for the development of effective anti-jamming communication strategies \cite{b10}.


Wireless jamming presents a considerable threat due to its covert and abrupt characteristics \cite{b11,b12,b13}, which can typically be classified into two distinct types: suppression jamming and dexterous jamming, according to the employed attack methods \cite{b14,b15}.  Suppression jamming involves the degradation of signal-to-noise ratio (SNR) of the received signal.   Conversely, dexterous jamming interferes with  specific operations at the receiver, including synchronisation and channel estimation.  Targeting these critical functions allows dexterous jamming to significantly impair the overall performance of wireless  systems.

The objective of jamming identification  is to determine the specific type of jamming signals \cite{b16,b17}.
The relationship between jamming identification accuracy and  performance metrics, such as bit-error rate and outage probability, is crucial for low-altitude wireless networks with various drone operations.  In dynamic environments where drones encounter advanced electromagnetic jamming threats, accurate jamming identification is essential for implementing the effective countermeasures.

Modern  jamming identification techniques can be categorised into two main paradigms: traditional feature-based approach and its deep learning counterpart.  The conventional methodology primarily utilizes artificial feature extraction  through a sequential processing including manual engineering of signal characteristics followed by pattern classification   \cite{b18,b19,b20}.
Feature extraction for jamming signals generally involves comprehensive analysis across multiple domains including time domain, frequency domain, and additional pertinent  domains.  Extracted features can be categorized into physical, spatial, statistical, and other characteristics.  Multi-domain feature fusion enables comprehensive signal characterization, significantly enhancing the discrimination capability of jamming signals. 
Feature selection criteria prioritize parameters with both physical interpretability and strong discriminative properties.  Commonly adopted features include higher-order cumulants for nonlinear signal analysis, peak-to-average ratio (PAPR), and spectral kurtosis for non-Gaussian component detection.
Jamming identification utilising artificial feature extraction generally employs classifiers to differentiate  various types of jamming signals, where  different classifiers demonstrate distinct degrees of complexity and performance in classification tasks.  Common classifiers include decision tree (DT) \cite{b21}, support vector machine (SVM) \cite{b22}, and artificial neural network (ANN).

Traditional jamming identification methodologies predominantly employ manual feature extraction from jamming signals, where the selection of discriminative features has prominent impact on classification performance.  While these  approaches offer certain advantages, e.g.  low implementation complexity,  they demonstrate fundamental limitations in modern electronic warfare scenarios. Firstly, these methods exhibit poor adaptability to dynamic jamming scenarios with evolving threat characteristics. Secondly, the decoupling of  feature engineering and pattern recognition  prevents the end-to-end performance optimization, which limits the jamming identification accuracy in complex electromagnetic environments. 

Deep learning has emerged as a transformative technology, significantly impacting application scenarios such as computer vision, natural language processing, and semantic segmentation \cite{b23,b24,b25,b26}.  Its remarkable capacity to extract and represent nonlinear relationships has significantly contributed to its effectiveness across multiple engineering fields \cite{b27,b28,b29}, such as millimetre wave beam management \cite{b30}, resource allocation \cite{b31}, and signal classification \cite{b32,b33,b34}.
Jamming identification adopting deep learning has drawn much attention recently, which removes the tedious task of manual feature extraction. 
Current approaches predominantly utilize convolutional neural networks (CNNs)  \cite{b35,b36,b37}, due to their validated efficacy in local feature extraction and automated parameter optimization through backpropagation. However, the inherent locality of CNN operations imposes fundamental limitations on network optimization, which has motivated the exploration of alternative architectures with enhanced modeling capabilities. Recently,  transformer architectures from natural language processing have been adopted for  jamming identification \cite{b10,b38}, which offer significant advantages through their global attention mechanisms.  Building upon these architectural advancements, recent research has increasingly focused on the security aspects of jamming identification systems adopting deep learning. Various methods have been developed to improve the resilience of neural networks against adversarial attacks \cite{b39,b40,b41,b42,b43,b44,b45}.   

While these approaches demonstrate promising capabilities, additional potential need to be carefully developed for enhancing network accuracy and robustness. This paper presents innovative transformer-empowered robust training methodologies aimed at the optimal trade-off between  accuracy and robustness in order to improve the security significantly for low-altitude wireless networks.   We propose a differential transformer framework for wireless jamming identification which could enhance the prediction accuracy when compared with the conventional transformer networks.  Additionally,  randomised masking training and consistent training  are developed to improve the  robustness against adversarial samples, which  allow the model to maintain high accuracy on clean data and possess effective defence mechanisms against adversarial attacks.  Our contributions are summarised as follows.
\begin{enumerate}[i.]
\item We propose  a differential transformer network for wireless jamming identification.  The traditional transformer network employs a self-attention mechanism for global feature extraction, which is prone to attention noise, resulting in disproportionate attention being assigned to irrelevant input regions.  The differential transformer network could address this limitation by employing differential computation of self-attention scores, thereby effectively reducing the attention noise.
\item   We propose a randomised masking training strategy to improve the robustness of transformer-based networks against adversarial attacks, which utilises multiple parallel feature extraction branches, each functioning with independently generated random masks. The occlusion pattern of each branch selectively obscures different regions of the input, which could enhance the prediction robustness by mitigating the influence of adversarial samples on individual branches and integrating outputs from multiple branches.
\item   We propose a consistent training strategy to enhance robustness against adversarial attacks,  which consists of two parallel pathways: one that is adversarially trained using randomised masking and noise perturbation, and another that is adversarial-agnostic, processing clean samples without any defence mechanisms.  Consistent training involves enforcing feature alignment between intermediate representations and regularising probability outputs for identical inputs, which facilitates knowledge transfer from the secure branch, providing the clean pathway with inherited adversarial resilience.
\item  Extensive simulations validate the benefits of the proposed differential transformer for jamming identification and the resilience of the two adversarial training strategies in the presence of adversarial samples.
\end{enumerate}

The paper is organized as follows. Section II outlines the signal model and the mathematical background. Section III introduces the novel differential transformer  framework and two  different defense training strategies for  jamming identification. In Section IV,  comprehensive  simulations are presented, demonstrating the efficacy of the proposed methodology in jamming identification. The  conclusion is drawn in Section V. 

\section{System Model}

In this section, we present the wireless jamming scenario and reception model of the jamming signal. We analyze the time-frequency characteristics of the received signals and transform them into time-frequency maps using the short-time Fourier transform. Finally, we describe the modeling of adversarial samples.\par
 
\subsection{Wireless Jamming}
\begin{figure}
	\centering
	\includegraphics[width=3.5 in]{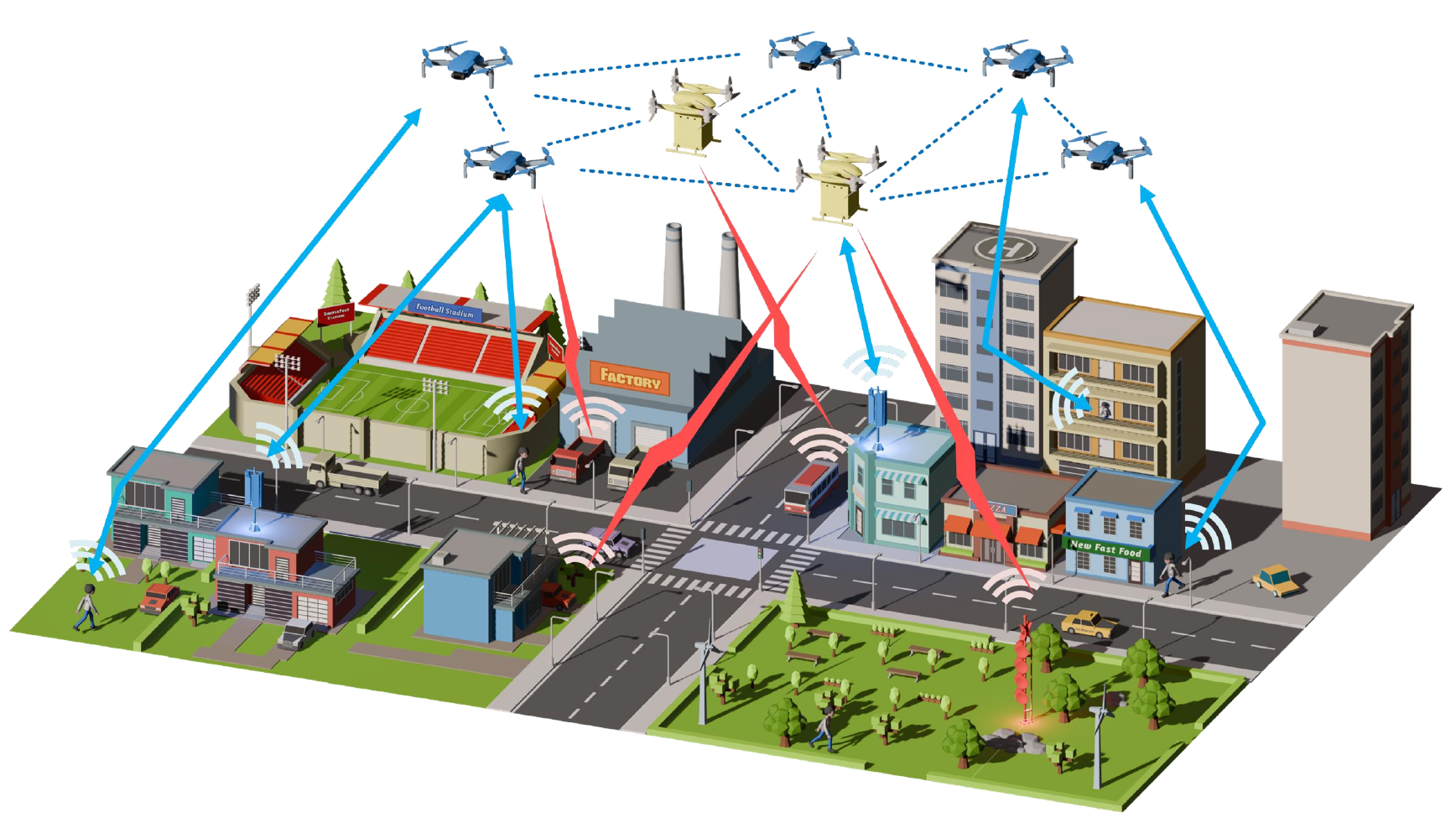}
	\caption{Wireless jamming scenario.}
	\label{fig0}
\end{figure}

Fig. \ref{fig0} depicts a jamming scenario in low-altitude wireless networks.  The red line denotes electromagnetic jamming signals produced by non-cooperative entities, specifically intended to interfere with drone communications.  Malicious interference sources substantially diminish communication quality.  Jamming signals generally originate from ground-based or airborne equipment, propagate through direct paths, and demonstrate significant jamming power.  High-power jamming significantly affects drone communication.
\subsection{Signal Model}
The reception model of the jamming signal is mathematically modeled as follows
\begin{equation}
	J(t) = z(t){e^{{\rm{j}}(2\pi {f_z}{\kern 1pt} t + {\varphi}(t))}}*h(t) + o(t){e^{{\rm{j}}(2\pi {f_o}{\kern 1pt} t + \phi (t))}}*g(t) + n(t){\kern 1pt},
\end{equation}
where $ J(t)$ denotes the received signal. $z\left( t \right)$ and $o\left( t \right)$ represent the jamming signals sent by non-cooperators and communication signals sent by cooperators, respectively. $h\left( t \right)$ and $g\left( t \right)$ denote  the unit channel impulse response of the interfering signal and the communication signal.  ${f_z}$ and  $\varphi \left( t \right)$ indicates the carrier frequency and initial phase of the interfering signal.  Similarly, ${f_o}$ and  $\phi \left( t \right)$   represent  the carrier frequency and initial phase of the communication signal. $\varphi \left( t \right)$ and $\phi \left( t \right)$ are uniformly distributed between $0$ and $2 \pi$. $n(t)$ indicates additive Gaussian white noise (AWGN).   The set of jamming candidates is defined as ${\bf{\Xi }} = \left\{ {{z_i}} \right\}_{i = 1}^M$, where ${z_m}$ denotes the $m$ type of jamming, and $M$ is the total number of jamming signal types. 

Let $P\big( z_i |J(t)\big)$ represent the conditional probability distribution of  $z_i$ given the observed signal $J(t)$. The final  determination of the category index $i^{\star}$  can be formulated as
\begin{equation}\label{eqMAP} 
	i^{\star} = \arg \max\limits_{i\in \{ 1,2,\cdots ,M \}} P\big(z_i | J(t)\big).
\end{equation}

We model jamming identification as a classification problem, thereby leveraging the powerful capabilities of neural networks for such tasks. Since neural networks typically take images as input, we transform the received signal into an image using time-frequency maps.
The time-frequency map  $X_T$ is derived using a short-time Fourier transform (STFT), which can be expressed as follows
\begin{equation}\label{eqSTFT} 
	X_T = \left| \sum\limits_{n = 0}^{N_{\rm{STFT}} - 1} J(n) w^{\text{H}}(n - n')\, e^{ -\textsf{j} \frac{2 \pi  k n}{N_{\rm{STFT}}} } \right|^2 ,
\end{equation}
where $J(n)$ represents the sampled value of the received signal $J(t)$. $w\left( n \right)$ and $w^{\text{H}}\left( n \right)$ are the window function and  its complex conjugate, respectively.  ${N_{{\rm{STFT}}}}$ represents the number of STFT points.   $n'$ and $k$ denote the discrete indices of time and frequency, respectively. The set of jamming candidates involves eight typical jamming signals, including continuous wave (CW), linear frequency modulation (LFM), amplitude modulation (AM),  triangular  frequency modulation (TFM), binary phase shift keying (BPSK),  noise amplitude modulation (NAM), quadratic frequency modulation (QFM) and  sinusoid frequency modulation (SFM). 

\subsection{Adversarial Attacks}
We examine deep learning models that maintain robustness against adversarial attacks while preserving the prediction accuracy of untainted samples.  The objective of the adversarial attack is to induce erroneous predictions in the deep learning network.  The adversarial attack signal is represented by $\delta$, and an adversarial sample is characterised as $x_a=x+\delta$. In this context, the output of the network for an input of $x_a$ diverges from the true label $y$.  To enhance the stealthiness of the constructed adversarial sample, we impose a restriction on the p-norm of the adversarial attack signal $\delta$, limiting the perturbation to a maximum value of $ \mathrm{P}_{\max }$, formulated as
\begin{equation}
	\begin{aligned}
		& \arg \max _{\delta} \operatorname(f(x_a; \theta) \neq f(x; \theta)), \\
		& \text { s.t. }\left|\delta\right|_p \leq \mathrm{P}_{\max },
	\end{aligned}
\end{equation}
where $f(\cdot; \theta)$ represents neural network with parameter $\theta$.  Since the goal of adversarial attacks is to maximize the destruction of the network's functionality, we do not restrict the type of attack. This implies that the adversary  possesses prior information, such as the weights and structures of deployed networks. However, the prediction model lacks any prior information about the counterattack.

\section{ Proposed Methodology}
\label{sec:guidelines}
This section introduces a differential transformer network designed for jamming identification, effectively eliminating the attention noise associated with traditional transformers.  This paper proposes a randomised masking training method and a consistent training method to enhance the network's robustness against adversarial attacks.

\subsection{Differential Transformer}
The primary advantage of the transformer network lies in its global extraction property. This attribute stems from the core computational unit of the transformer, known as self-attention.  First, we start with a brief introduction to the self-attention operation in traditional transformers \cite{b47}.

Let $X \in {\mathbb{R}^{N \times C}}$ denote the input, where $N$ denotes  the length of the input sequence, and $C$ denotes the dimension. When computing self-attention,  we first  obtain the three matrices of queries $Q \in {\mathbb{R}^{N \times d}}$, keys $K \in {\mathbb{R}^{N \times d}}$ and values $V \in {\mathbb{R}^{N \times d}}$, which can be expressed as $Q=XQ_W$, $K=XK_W$, $V=XV_W$, respectively,  where $Q_W \in {\mathbb{R}^{C \times d}}$, $K_W \in {\mathbb{R}^{C \times d}}$ and $V_W \in {\mathbb{R}^{C \times d}}$ are learnable matrices, and $d$ denotes the dimension.  

The attention score in self-attention can be expressed as
\begin{equation}
	Z = \operatorname{softmax}\left(\frac{Q K^T}{\sqrt{d}}\right) .
\end{equation}

The output $Z\in {\mathbb{R}^{N \times N}}$ represents the interrelationships between the elements of the input, and the final result of self-attention is obtained by multiplying $Z$ by $V$, formulated as
\begin{equation}
	S = \text { Attention }(Q, K, V)=Z V,
\end{equation}
where $S\in {\mathbb{R}^{N \times d}}$ represents output of  self-attention layer.  

A recent study \cite{b46} indicates that transformers compute the attention score $Z$ incorporating attention noise.  This describes the phenomenon in which attention scores exhibit larger values in non-informative segments of input $X$, while critical segments receive comparatively lower attention scores.   We present the differential transformer network, which effectively mitigates attention noise.  The fundamental concept of the differential transformer involves the creation of a pair of attention scores and the formulation of a differential representation of these scores.

Next, we will elaborate on the differential transformer structure. First, we expand the channel dimensions of $Q_W$ and $K_W$ to twice their original size, namely $Q_W \in {\mathbb{R}^{C \times 2d}}$, $K_W \in {\mathbb{R}^{C \times 2d}}$. The resulting channel dimension of the $Q\in {\mathbb{R}^{N \times 2d}}$ is also twice as large, and it is divided by $Q_1\in {\mathbb{R}^{N \times d}}$ and $Q_2\in {\mathbb{R}^{N \times d}}$ along channel dimension, formulated as 
\begin{equation}
\left[Q_1 ; Q_2\right]=X W_Q.
\end{equation}

Similar to $Q$, $K_1\in {\mathbb{R}^{N \times d}}$ and $K_2\in {\mathbb{R}^{N \times d}}$ can be expressed as
\begin{equation}
	\left[K_1 ; K_2\right]=X W_K.
\end{equation}

\begin{figure}
	\centering
	\includegraphics[width=2.8 in]{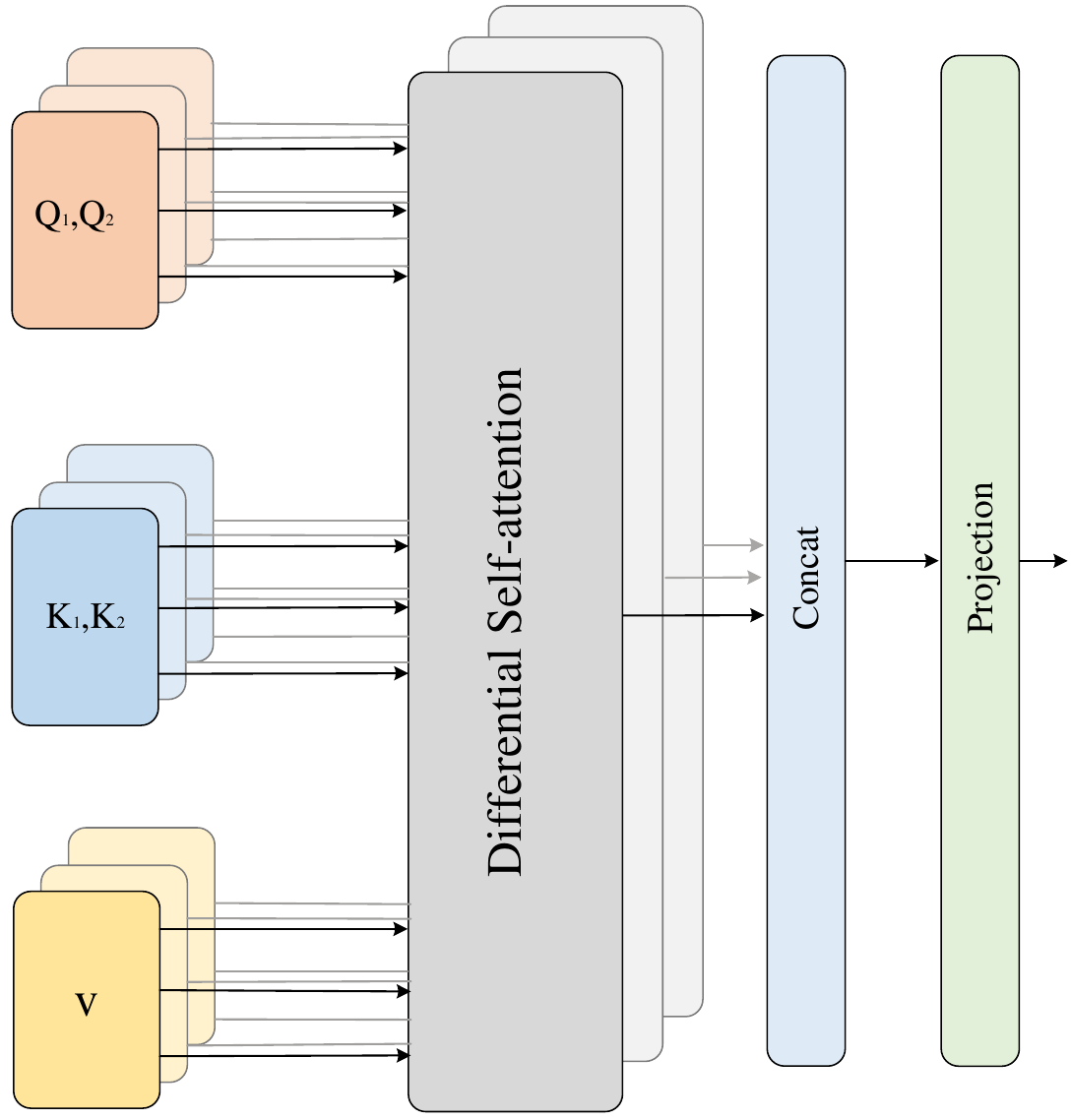}
	\caption{Differential self-attention structure.}
	\label{fig1}
\end{figure}
The differential self-attention can be represented as
\begin{equation}
	\begin{aligned}
		\operatorname{Diff}(X)=\left(\operatorname{softmax}\left(\frac{Q_1 K_1^T}{\sqrt{d}}\right)-\lambda \operatorname{softmax}\left(\frac{Q_2 K_2^T}{\sqrt{d}}\right)\right) V,
	\end{aligned}
\end{equation}
where $\lambda$ represents hyperparameter and is set to 0.8 in this paper. The differential self-attention is shown in Fig. \ref{fig1}.

\begin{figure*}
	\centering
	\includegraphics[width=6.1 in]{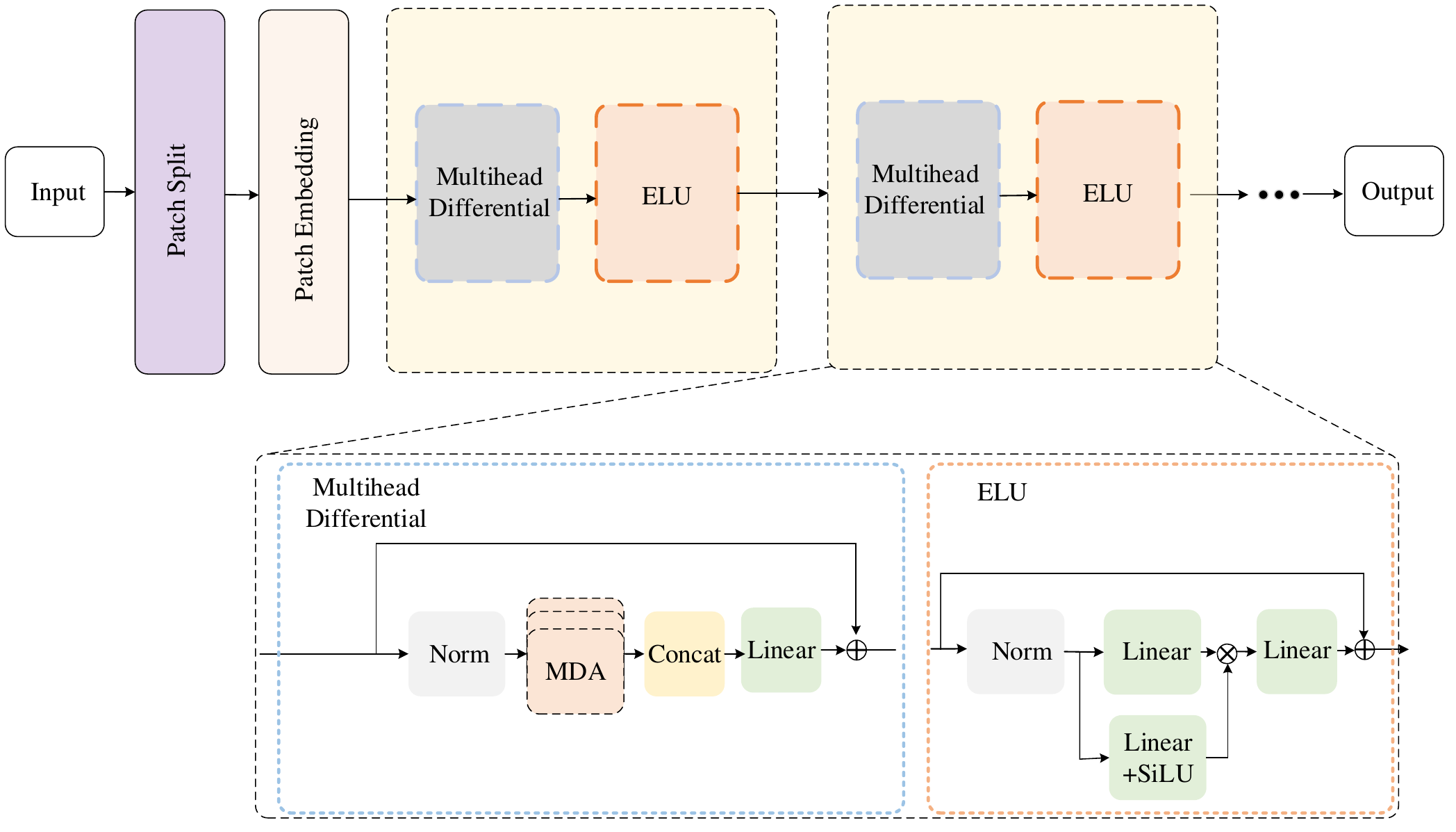}
	\caption{Feature extraction with differential transformer  begins by partitioning the input into patches. Following patch embedding, the resulting features go  through a multi-head difference layer to capture the global representations. An ELU activation function is subsequently applied to strengthen the inter-channel feature interactions.}
	\label{fig2}
\end{figure*}

Similar to transformer's multi-head self-attention layer, a multi-head differential attention  can be constructed to increase the diversity of extracted features. 
The number of multiple heads is set to $h=C/d$, with each head corresponding to different learnable matrices $W_Q^i$, $W_K^i$, $W_V^i$, and $i=1,2,...,h$. We concatenate the different attention heads into a single matrix and then map this matrix to get the final output,  formulated as
\begin{equation}
	\begin{aligned}
		\operatorname{head}_i & =\operatorname{Diff}\left(X ; W_Q^i, W_K^i, W_V^i, \lambda\right), \\
		\overline{\operatorname{head}_i} & =\left(1-\lambda_{\text { }}\right) \cdot \operatorname{LN}\left(\operatorname{head}_i\right), \\
		\operatorname{MultiDiff}(X) & =\operatorname{Concat}\left(\overline{\text { head }_1}, \cdots, \overline{\text { head }_h}\right) W^O,
	\end{aligned}
\end{equation}
where  $W^O  \in {\mathbb{R}^{C \times C}} $ is learnable projection.   $\operatorname{LN}\left(\cdot\right)$ represents the layer normalization operation.

Following the multi-head differential attention operation, the network employs a enhanced linear unit (ELU) structure to facilitate channel feature interaction, further enhancing the network's feature extraction capabilities.
The ELU, which consists of three layers of fully connected (FC) layer, can be expressed as
\begin{equation}\label{eqMLP} 
	\text{ELU}\big(\alpha\big) = \big(\omega \left( \alpha W_1\big) \otimes \alpha W_2\right)\big) W_3 ,
\end{equation}
where $W_1\! \in\! \mathbb{R}^{C \times (b\cdot C)}$, $W_2\! \in\! \mathbb{R}^{C \times (b\cdot C)}$  and $W_3\! \in\! \mathbb{R}^{C \times C}$ correspond to the weights of three FC, respectively, where $b$ is referred to as the expanding ratio. $\otimes$ represents Hadamard product. The activation function $\omega$ is chosen to be the  sigmoid linear unit (SiLU).

Similar to the transformer architecture, we incorporate a residual connection and layer normalization operation, which is denoted as follows
\begin{equation}
	\begin{array}{l}
		{O} = {\rm{MultiDiff}}\left( {{\rm{NL}}\left( {{X}} \right)} \right)  + {X},\\
		{O}' = {\rm{ELU}}\left( {{\rm{NL}}\left( {O} \right)} \right) + {O},\\
	\end{array}
\end{equation}
where ${O} \in {\mathbb{R}^{N \times {C}}}$ and ${{O}'}  \in {\mathbb{R}^{N  \times C}} $ are the output of multihead differential attention and ELU. The overall network structure is shown in Fig. \ref{fig2}, which serves as a feature extractor for interfering signals. The patch split and patch embedding operations are analogous to those used in a traditional transformer.

\subsection{Randomized Masking Training }
\begin{figure*}
	\centering
	\includegraphics[width=6.3 in]{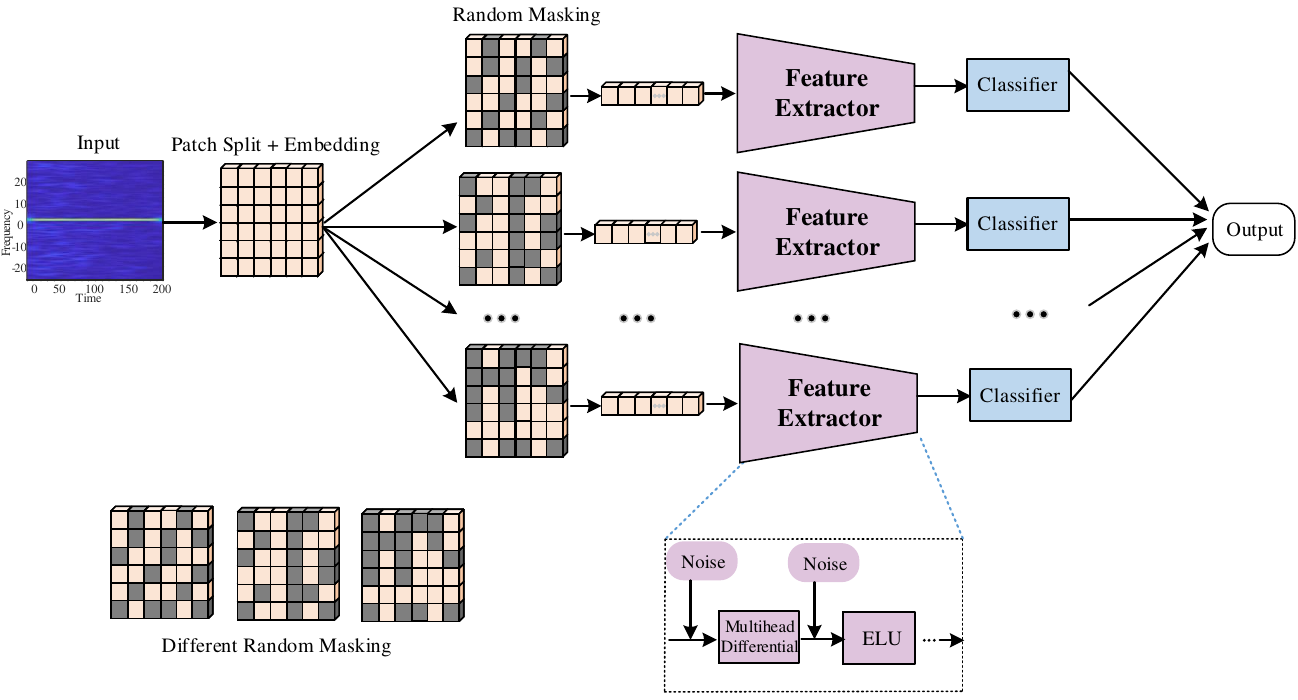}
	\caption{Randomized masking training for improving model robustness.  }
	\label{fig3}
\end{figure*}

In this subsection, we propose a randomized masking training to  defense against attacks, as shown in Fig. \ref{fig3}. 

Before detailing the methodology, we will elaborate on its underlying insights. When utilizing transformer-based networks to extract features from an input, the input signal is divided into multiple non-overlapping patches. In this paper, when deploying the differential transformer-based network for feature extraction, the final output is obtained by performing differential self-attention and ELU operations on these patches.

Upon contamination of the input signal by adversarial perturbations, our defence mechanism activates a strategic decomposition process.  In accordance with the operational principles of differential transformers, the contaminated input is initially divided into several non-overlapping patches.  This patch-wise decomposition disperses adversarial contamination unevenly across various spatial regions, resulting in a heterogeneous distribution of attack potency, with some patches exhibiting significant contamination while others remain relatively unaffected.   To mitigate this adversarial influence, we propose the selection of a subset of patches as inputs (masking partial patches) to the network on each occasion, thereby diminishing the efficacy of the adversarial attack.  By dynamically masking a subset of patches during each forward pass, we reduce the attack's influence on the model's decision-making process.  The highest resistance to adversarial attacks is attained when the subset of input patches excludes all heavily contaminated patches.  This markedly diminishes the effect of the attack strategy on the classifier.  In conclusion, various patch masking methods can be utilised to reduce the effects of adversarial attacks. 
The sequence derived from patch splitting and patch embedding of the input is represented as $X_p\in {\mathbb{R}^{N \times C}}$, where $N$ signifies the number of patches and $C$ indicates the dimension of each patch.   We develop $k$ parallel branches that utilise shared parameters, implementing different masking strategies for each branch.  Let $S_i$ ($i=1,2,...,k$) represent any of the potential masking strategies whereby $X_p$ disregards the outcomes of partial patches.   Each branch consists of three components: (a) a base classifier, (b) feature extraction utilising multi-head differential attention and ELU modules, and (c) randomised masking employed to select a subset of available patches for feature extraction.  The final output of the network is represented as the aggregate of predictions from $k$ classifiers, expressed as
\begin{equation}
p = \sum\limits_{i = 1}^k {g_i\left( {\phi_i \left( {{X_p}\left| {{S_i}} \right.} \right)} \right)} ,
\end{equation}
where $g_i(\cdot)$ and $\phi_i(\cdot)$ represent the $i$-th branch classifier and feature extraction, and $p$ denotes the output of the model. 

The idea behind this is that various masking strategies diminish the effectiveness of patches under adversarial conditions.  This indicates that certain adversarial patches significantly affecting the classifier are obscured, enabling the network to demonstrate robustness.   Furthermore, Gaussian white noise has been incorporated into the input features of the multi-head differential attention and ELU module.  This modification reduces the effects of the adversarial signal.

The proposed strategy is appropriate for transformer-based network structures for two reasons:  Transformers, in contrast to CNNs, transform the input into a collection of patches.  This endows transformers with the capability to omit certain patches, thereby disregarding portions of the input, which aligns well with the masking strategy.  Transformers possess the ability to extract global features, enabling the exploration of connections among unobscured patches, the joint inference of details within these patches, and the reduction of masking effects.  Conversely, the convolutional kernels of CNNs encompass areas with a greater extent of masked regions, which notably influences the computation of these kernels.

Two randomised masking strategies are employed: continuous masking and discrete masking.  (a) Continuous masking: The length of the masking region is defined by a sequence of $n$, with the starting position randomly chosen to ensure that all patches within that range are obscured.
(b) Discrete masking: The quantity of masked patches is set at $n$.  Subsequently, $n$ patches are randomly selected from various positions to be masked, without necessitating that the masking be applied to consecutive patches.  Our simulations indicate that the performance of these two types of masking is similar.

Also, only these unmasked patches go through the transformer-based classifier, thus avoiding the redundant computation of all masked patches. We consider the dataset $D = \left( {{x_i},y_i} \right)_{i = 1}^U$, where $U$ refers to the number of samples, and $x_i$ and $y_i$ denote the $i$-th sample and  corresponding   label. The randomized masking training is given by Algorithm 1.

\subsection{Consistent Training }

The objective is to create a network that exhibits strong performance on both adversarial and clean samples.  This necessitates both resilience against adversarial attacks and elevated recognition accuracy on untainted samples.

We propose consistent learning schemes to ensure the network generates uniform outputs for both adversarial and non-adversarial versions of identical samples.  We impose a probabilistic consistency constraint on each sample and its adversarial counterpart.

Nonetheless, a significant challenge associated with this approach is the absence of adversarial samples.  Despite generating adversarial samples for consistent training, the trained network may remain susceptible to adversarial attacks utilising alternative construction strategies.

In response to the limitations posed by the absence of adversarial samples, we present our randomised masking training algorithm as a solution.  The proposed consistency learning seeks to reduce the impact of adversarial attacks, independent of exclusive reliance on supervisory information.

\begin{figure*}
	\centering
	\includegraphics[width=5.7 in]{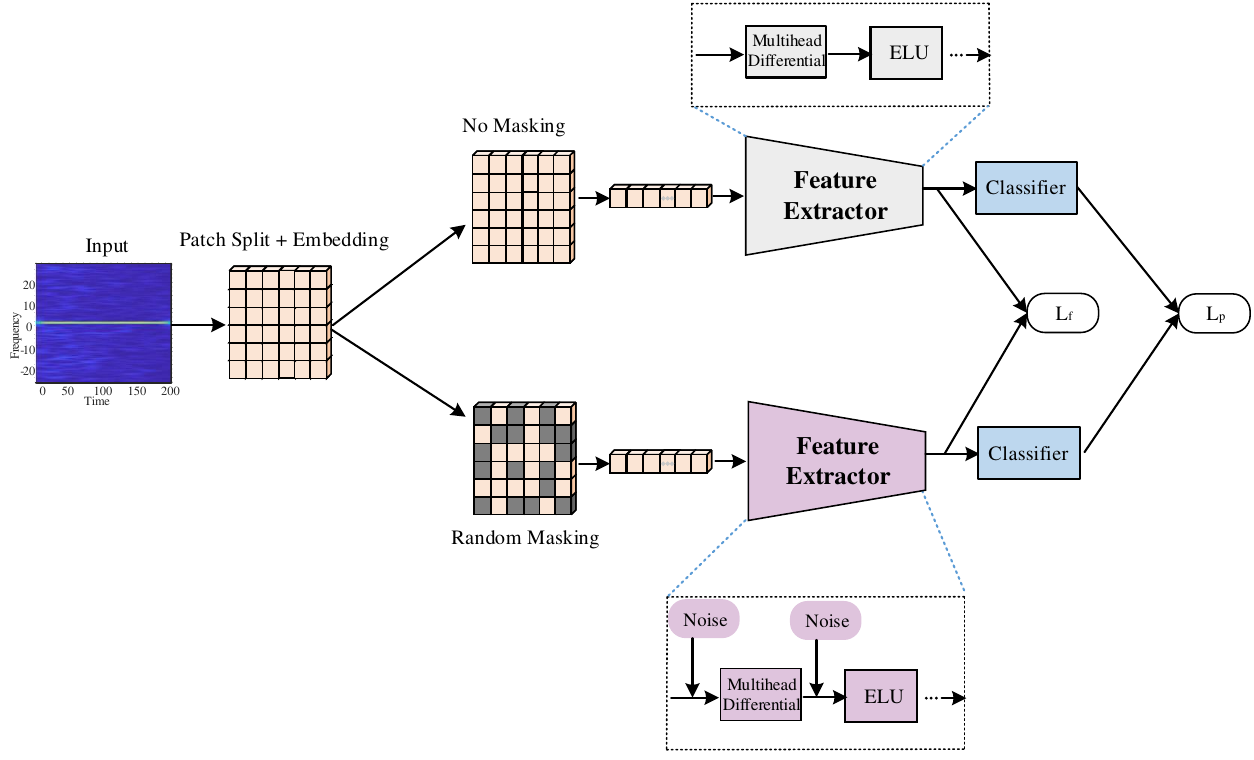}
	\caption{Consistent training methods to improve model robustness. The purple feature extractor handles features with artificially added noise, while gray feature extractor handles features without  artificially added noise. }
	\label{fig4}
\end{figure*}

We construct two network branches: a robust processing branch and a regular branch. The parameters of the two branches are shared, except that the robust processing branch leverages the robustness strategy outlined in randomized masking training.
 The input sample set is denoted as
$X_0=\left\{ {{x^i}} \right\}_{i = 1}^U$. We denote the intermediate features  both branches as:  $\left\{ {{Z_1},{Z_2}} \right\} = \left\{ {{z_1^i},{z_2^i}} \right\}_{i = 1}^U$, where $z_1=\phi \left( { x ;\theta } \right)$ and  $z_2=\phi \left( { x_m ;\theta } \right)$ denote feature representation of  two branches for the sample $x$, where $\phi \left( { \cdot ;\theta } \right)$ represents the feature extractor with learnable parameters $\theta$, and $x_m$ represents processing of  randomized masking training. The parameters $\theta$ of the feature extractor will be optimized through an objective function designed to ensure that the features  $z_1$ and $z_2$ are closely aligned.  We refer to the representation of $z_1$ and $z_2$ from the same distribution, also known as a positive sample pair.

Our feature extractor $\phi \left( { \cdot ;\theta } \right)$ is trained to produce proximity characteristics for positive sample pairs. Thus, the loss function can be expressed as
\begin{equation}
	{\mathcal{L}_f} = {\sum\limits_{i = 1}^U {\left| {z_1^i - z_2^i} \right|} ^2}.
\end{equation}

We denote the  final probabilities of both treatments as:  $\left\{ {{P_1},{P_2}} \right\} = \left\{ {{p_1^i},{p_2^i}} \right\}_{i = 1}^U$, where $p_1$ and  $p_2$ denote output probabilities of two branches for the sample $x$. The final probabilities can be represented by $p_1=g \left( { z_1 ;\gamma } \right)$ and $p_2=g \left( { z_2 ;\gamma } \right)$, where $g \left( { \cdot ;\gamma } \right)$ denotes the classifier with parameters $\gamma$. 

At the same time, we approximate the distance between $p_1$ and  $p_2$ in a similar way, denoted as
\begin{equation}
	\mathcal{L}_p = {\sum\limits_{i = 1}^U {\left| {p_1^i - p_2^i} \right|} ^2}.
\end{equation}

The final overall loss can be expressed as
\begin{equation}
	\mathcal{L}_e = \mathcal{L}_c+{\beta _1}{\mathcal{L}_f} + {\beta _2}{\mathcal{L}_p},
\end{equation}
where $\mathcal{L}_c$ represents cross entropy loss function, and ${\beta _1}$ and  ${\beta _2}$ indicate hyperparameters. ${\beta _1}$  and ${\beta _2}$ are set to 0.2 in this paper.  Evidently, this formula allows the network to capture the interplay between the two processing methods, rendering it less susceptible to minor input variations and yielding more robust predictions.

Upon completion of training, the regular branch inherits the robustness of the other branch. Consequently, we discard the robust processing branch and solely employ the regular branch for prediction. Algorithm~\ref{alg:AG} outlines the training and testing procedure for the proposed consistent training.

\begin{algorithm}[ht]
\caption{The  process of the proposed randomized masking training.}
\label{alg:AG}
\LinesNumbered
\KwIn{Training data $D = \left( {{x_i},y_i} \right)_{i = 1}^U$,  training epoches $T_e$, $K$ mini-batches in each epoch, the learning rate $\eta$, masking rate $m_r$, number of branches $k$}
\KwOut{Trained model.}
Construct $k$ parallel differential transformer branches;\\
\For{$k = 1, 2, \cdots, T_e$}{
\For{$j = 1, 2, \cdots, K$}{
Introduce a randomized masking strategy $S_i$ with occlusion rate $m_r$ for each branch and artificially add noise to each branch;\\
Calculate the result $p$ by Eq. (13);\\
Update parameters  by  cross-entropy  via an end-to-end approach with a  learning rate $\eta$;
}
}

Save the  model;

\end{algorithm}

\emph{}

\begin{algorithm}[ht]
	\caption{The  process of the proposed consistent training.}
	\label{alg:AG}
	\LinesNumbered
	\KwIn{Training data $D = \left( {{x_i},y_i} \right)_{i = 1}^U$,  training epoches $T_e$, $K$ mini-batches in each epoch, the learning rate $\eta$, masking rate $m_r$}
	\KwOut{Trained model.}
	Construct two parallel differential transformer branches, namely robust branch and regular branch;\\
	\For{$k = 1, 2, \cdots, T_e$}{
		\For{$j = 1, 2, \cdots, K$}{
			Introduce a randomized masking strategy $S$ with masking rate $m_r$ and artificially add noise  for robust branch;\\
			Calculate the output of both branches;\\
			Update parameters  by  (16)   via an end-to-end approach with a  learning rate $\eta$;
		}
	}
		
	Save the  regular branch;
\end{algorithm}

\section{Simulation Results}
This section provides an evaluation of the proposed algorithm.  Subsection A presents the experimental dataset and the conditions for training.  Subsection B outlines the structure of the network model utilised in this study.  Subsection C evaluates the prediction performance and robustness of the proposed algorithm via simulations.  The simulation results indicate that the two proposed robustness training methods improve the model's resilience to adversarial samples.

\subsection{Simulation Setup}
We construct  a jamming dataset comprising eight jamming types: AM, BPSK, CW, LFM, NAM, QFM, SFM, and TFM. To simulate real-world jamming signals, the frequency and bandwidth of each jamming type are randomly varied, as shown in Table \ref{tab1}. The communication signal is an orthogonal frequency division multiplexing (OFDM) signal with a subcarrier spacing of 15 kHz and 1200 subcarriers, centered at 0 Hz. The jamming signal and the OFDM signal are transmitted over single-path Rice and multipath Rayleigh channels, respectively. The Rice factor is 15 dB, while the delay and gain of the multipath Rayleigh channel were set to [0 1 2 3 4 5] $\times 10^{-7}$ s and [0 --4 --8 --12 --16 --20] dB, respectively.

\begin{table}[]
	\centering
	\caption{Parameters of the interference signals.}
	\label{tab1}
	\begin{tabular}{ccccc}
		\hline
		Interference patterns                  & Parameters      & Value Range                            \\
		\hline
		\multirow{2}{*}{CW}   & $f_c$(MHz) & from --25 to 25                       \\
		& ISNR(dB) & from --14 to 8                    \\
		\hline
		&  $f_c$(MHz) & from --25 to 25                    \\
		\multirow{2}{*}{SFM}                       & Bandwidth(MHz) & from 10 to 50                      \\
		& ISNR(dB) &  from --14 to 8                      \\
		& Period  $(ms)$ &  0.01 to 0.1                   \\
		\hline
		&   $f_c$(MHz) & from --25 to 25                        \\
		\multirow{2}{*}{QFM}                       & Bandwidth(MHz) & from 10 to 50                      \\
		& ISNR(dB) &  from --14 to 8                      \\
		& Period  $(ms)$ &  0.01 to 0.1                 \\
		\hline
		&  $f_c$(MHz) & from --25 to 25                      \\
		\multirow{2}{*}{TFM}                      & Bandwidth(MHz) & from 10 to 50                      \\
		& ISNR(dB) &  from --14 to 8                      \\
		& Period  $(ms)$ & 0.01 to 0.1                   \\
		\hline
		&  $f_c$(MHz) & from --25 to 25                        \\
		\multirow{2}{*}{BPSK}                       & Bandwidth(MHz) & from 1.5 to 5                      \\
		& ISNR(dB) &  from --25 to 25                      \\
		\hline
		&  $f_c$(MHz) & from --25 to 25                      \\
		\multirow{2}{*}{LFM}                       & Bandwidth(MHz) & from 10 to 50                      \\
		& ISNR(dB) &  from --14 to 8                      \\
		& Period  $(ms)$ &  0.01 to 0.1                   \\
		\hline
		&  $f_c$(MHz) & from --25 to 25                     \\
		\multirow{2}{*}{AM}   & Bandwidth(MHz) & from 1.5 to 5                      \\
		& ISNR(dB) &  from --14 to 8                      \\
		
		\hline
		&  $f_c$(MHz) & from --25 to 25                        \\
		\multirow{2}{*}{NAM}                       & Bandwidth(MHz) & from 1.5 to 5                      \\
		& ISNR(dB) &  from --14 to 8                      \\
		\hline
	\end{tabular}
\end{table}

The network processes the time-frequency representation of the jamming signal as input.  The dataset comprises 400 samples for each jamming type at every interference-to-signal-plus-noise ratio (ISNR), with a training to test set ratio of 3:1.  The ISNR varies from -14 dB to 8 dB.  The network employs the stochastic gradient descent (SGD) optimiser with a learning rate of 0.001 over 50 epochs.

\subsection{Model Structure}

The proposed network, illustrated in Table \ref{tab2}, utilises the time-frequency graph as input.  The time-frequency diagram is a 40 $\times$ 40, 3-channel RGB image.  The input is divided into 100 patches, each containing 48 channels.  Patch embeddings are utilised to extract initial features.  Two layers of differential attention and ELU layers are subsequently applied.  Global average pooling is employed to extract features, which are subsequently processed through a fully connected layer to produce the output.  The convolutional layer is represented as Conv(channel, kernel, stride), where channel, kernel, and stride refer to the channel dimensions, the size of the convolutional kernel, and the step size of the convolution, respectively.  MultiDiff(32, 4) denotes a multi-head differential attention mechanism characterised by 32 channels and 4 heads.  ELU(32) denotes an ELU layer comprising 32 channels, while FC(8) indicates a fully connected layer containing 8 neurones.  Batch normalisation (BN) and the rectified linear unit (ReLU) serve as the normalisation and activation functions, respectively.  GAP stands for global average pooling.  The overall network's size and scale can be modified through the addition or removal of differential attention layers.

\begin{table}[]
	\caption{ structure of Proposed method.}
	\label{tab2}
	\setlength{\tabcolsep}{3pt}
	\begin{center}
	\begin{tabular}{|c|c|}
		\hline
		Model           & Structure                              \\ \hline
		Input     & (3,40,40)    \\ \hline
		Patch split     & (100,48)    \\ \hline
		Patch embedding & Conv(32,3,1)+BN+ReLU                 \\ \hline
		Multi-head differential attention & MultiDiff(32,4)+ELU(32)       \\ \hline
		Multi-head differential attention & MultiDiff(32,4)+ELU(32)  \\ \hline
		Output     & GAP + FC(8)                                  \\ \hline
\end{tabular}
\end{center}
\end{table}

\subsection{Recognition Performance}

The experimental component consists of two parts: jamming identification with clean samples and with adversarial samples.  Initially, we assess the performance using clean samples.

To evaluate the performance advantages of our proposed network architecture, we conduct comparative experiments against five state-of-the-art jamming identification networks: CNN \cite{b23}, ResNet \cite{b24} representing deep residual learning, SEnet \cite{b26} with channel attention mechanisms, Transformer \cite{b47} for global attention modelling, and our proposed method.  The analysis centres on two essential dimensions of model performance: identification accuracy and computational efficiency. 

To ensure a rigorous and equitable evaluation, we uphold consistent experimental conditions across all methods being compared, encompassing the training dataset, testing scenario, input dimensions, and hardware platform.  The computational complexity is quantitatively assessed through floating point operations (FLOPs), with comprehensive comparisons provided in Table \ref{tab3}.  This metric offers a hardware-independent assessment of model efficiency, which is particularly crucial for applications involving real-time jamming recognition.

\begin{figure}
	\centering
	\includegraphics[width=3.5 in]{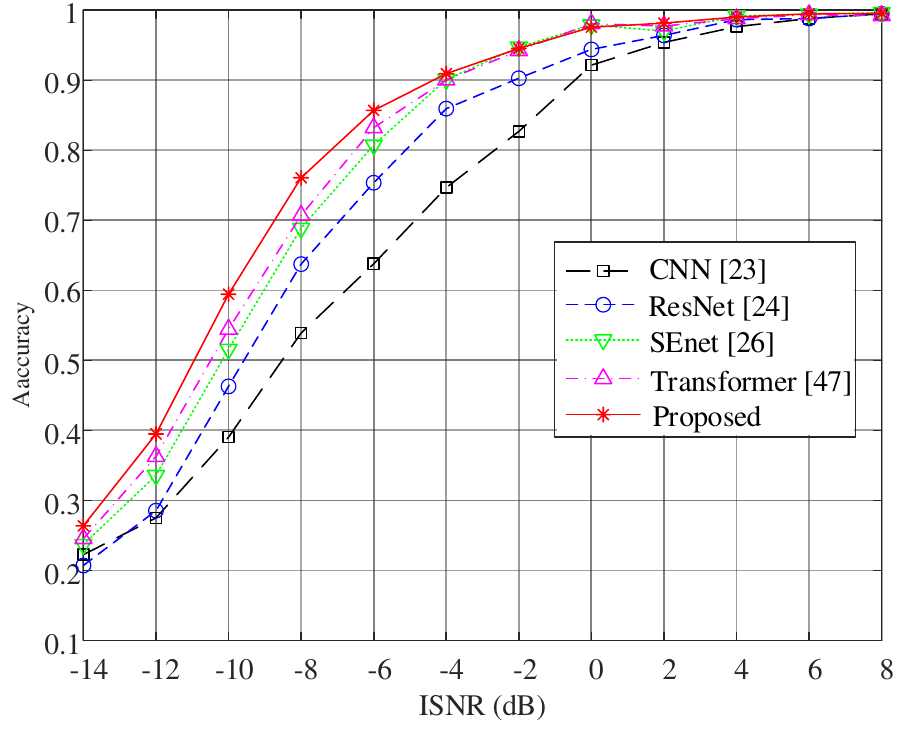}
	\caption{Recognition accuracy of different models.}
	\label{fig5}
\end{figure}

\begin{figure}
	\centering
	\includegraphics[width=3.5 in]{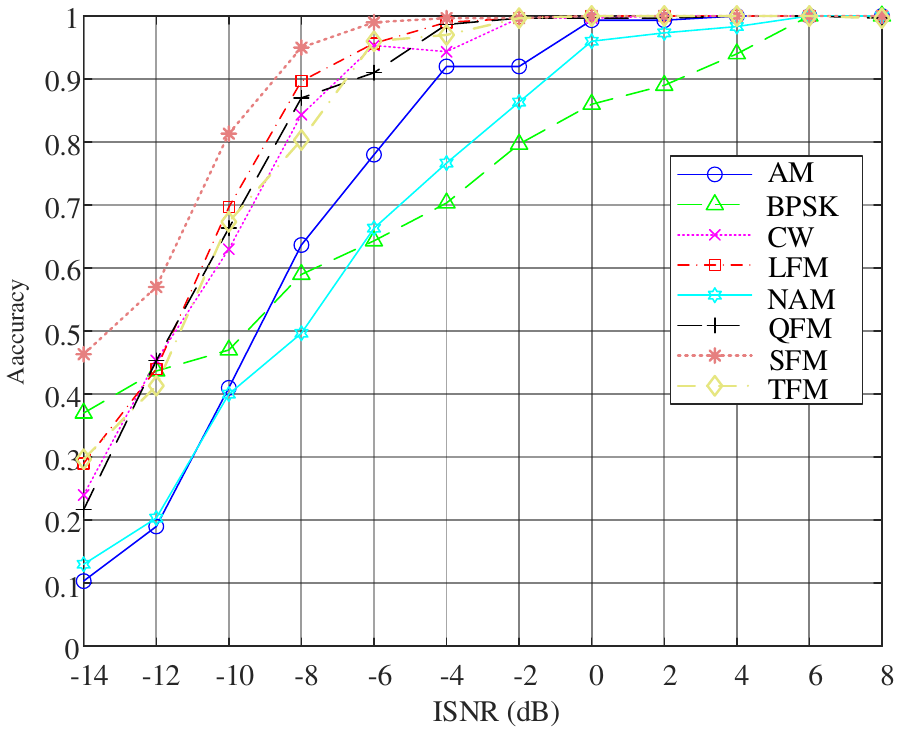}
	\caption{Recognition accuracy of the differential transformer model for each jamming class.}
	\label{fig6}
\end{figure}

As shown in Table \ref{tab3}, the FLOPs of the different models are comparable, ensuring a fair comparison. Notably, the proposed model has the lowest computational FLOPs of all the schemes. 

\begin{table}[htbp]
	\caption{ FLOPs  of different models.}
	\label{tab3}
	\setlength{\tabcolsep}{3pt}
	\begin{center}
		\begin{tabular}{|c|c|c|c|c|c|}
			\hline
			Model   & CNN   & ResNet  & SEnet & Transformer  & Proposed \\ \hline
			FLOPs($\times 10^6$) & 1.66 &  1.71 &   1.68     & 1.99       &    1.63      \\ \hline
		\end{tabular}
	\end{center}
\end{table}

Fig. \ref{fig5} demonstrates that the recognition accuracy of each model typically improves as ISNR increases.  At -14 dB, the recognition accuracy for all models falls below 30\%.  At 8 dB, the recognition accuracy of all models nears 100\%.  The proposed scheme consistently demonstrates superior recognition accuracy compared to other models across various ISNRs, highlighting its advantages.  The proposed method exhibits enhanced performance, achieving improvements of 5\%, 6\%, 10\%, and 21\% over the Transformer, SENet, ResNet, and CNN baselines, respectively, at -8 dB.  The proposed method demonstrates an effective enhancement in recognition performance through the introduction of a differential form of the attention mechanism. This approach mitigates attention noise, accurately captures critical discriminative regions of the interference signal, and subsequently improves prediction accuracy.

\begin{figure}[htbp]
	\centering
	\includegraphics[width=3.5 in]{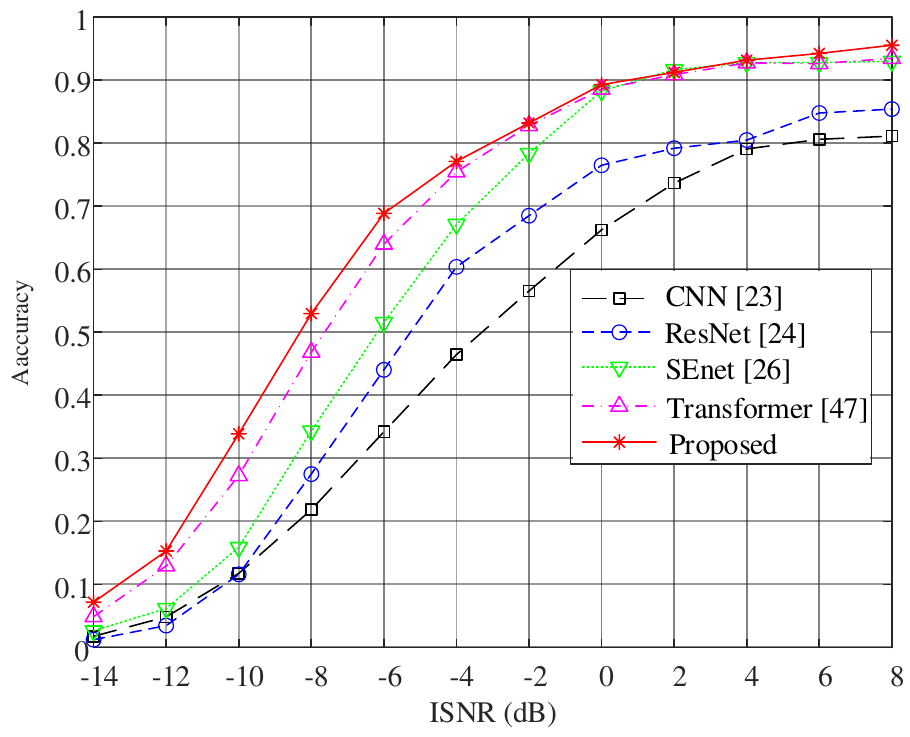}
	\caption{Recognition accuracy of each model under an attack perturbation of 3/255.}
	\label{fig7}
\end{figure}

\begin{figure}[htbp]
	\centering
	\includegraphics[width=3.5 in]{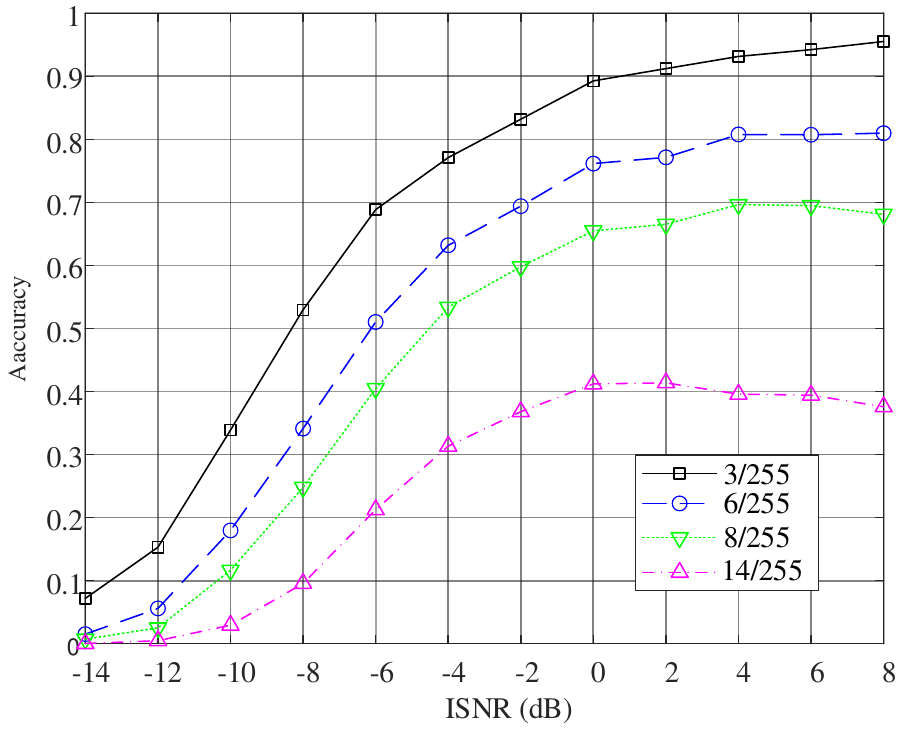}
	\caption{ Recognition accuracy of differential transformer under different perturbation factors.}
	\label{fig8}
\end{figure}

\begin{figure}[htbp]
	\centering
	\subfigure[Consistent training]{
		\includegraphics[width=3.5 in]{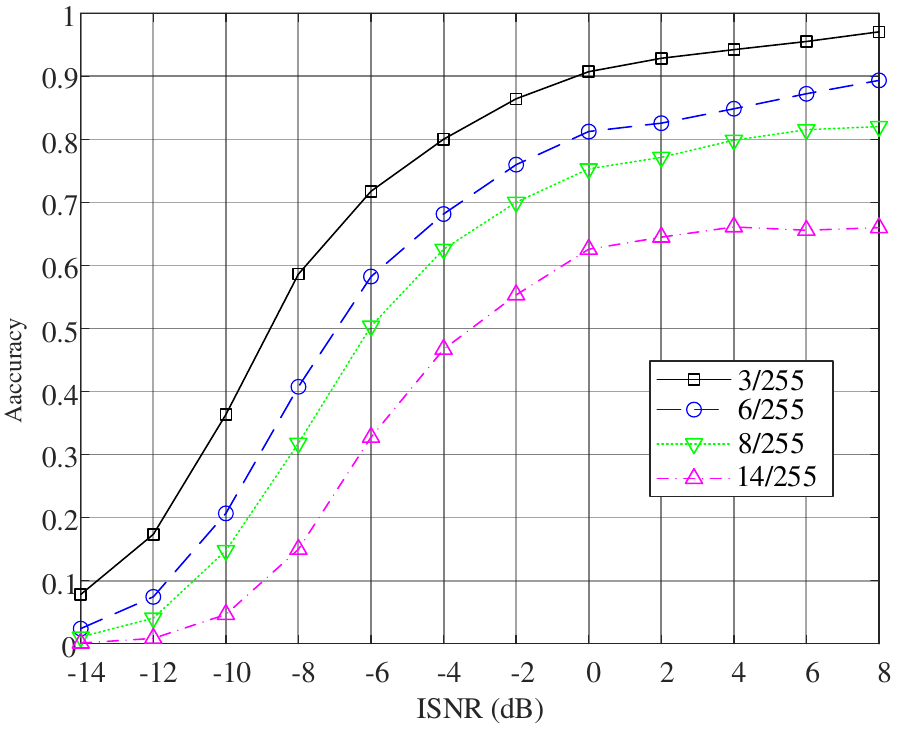}
		\label{fig9a}}
	\subfigure[Randomized masking training]{
		\includegraphics[width=3.5 in]{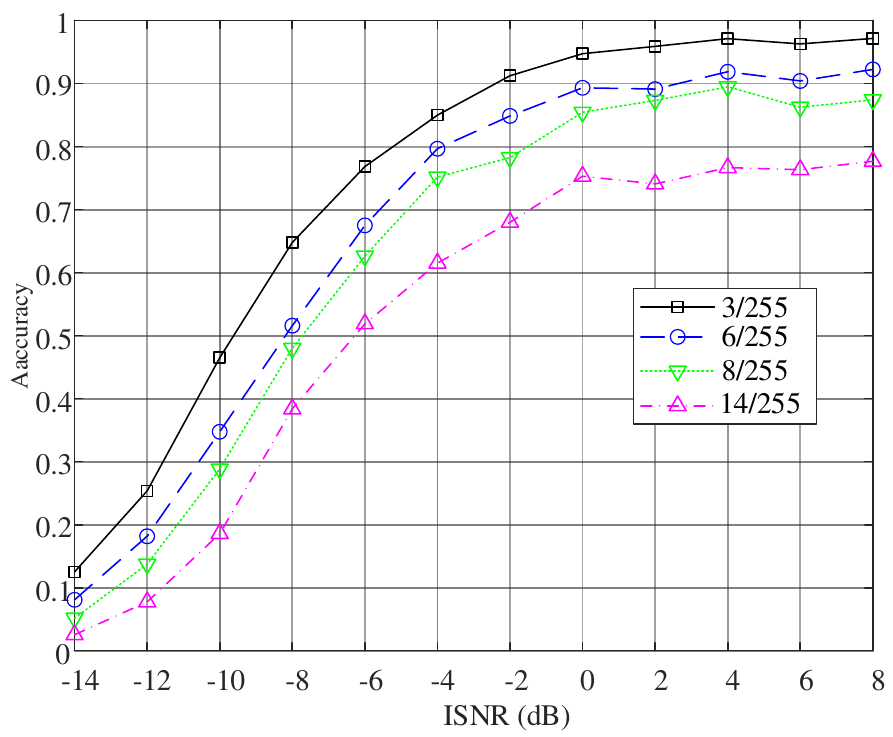}
		\label{fig9b}}%
	\caption{Defense effects under consistent training and randomized masking training with different perturbation factors.}
	\label{fig9}
\end{figure}

Fig. \ref{fig6} illustrates the recognition accuracy of the proposed scheme across different jamming styles at various ISNR levels.  The figure illustrates that with an increase in ISNR, the network's recognition accuracy for various jamming styles progressively nears 100\%.  The experimental results indicate a significant finding concerning the recognition performance of BPSK and NAM interference types.  The figure illustrates that these two interference signals demonstrate a markedly slower enhancement in recognition probability as ISNR increases, in contrast to other types of interference.  This observation is attributable to their fundamental signal characteristics.  The analysis of time-frequency characteristics indicates that BPSK and NAM interference demonstrate notable similarities in their spectrogram representations. 

A comprehensive robustness assessment of various models under adversarial conditions is conducted, as illustrated in Fig. \ref{fig7}.  The evaluation utilises the Fast Gradient Sign Method (FGSM) with a perturbation magnitude of 3/255, indicating a moderate-strength attack in a white-box context.  The analysis presents several significant findings.  (a) General Vulnerability Trends: All assessed models demonstrate notable performance decline when subjected to attacks.   The recognition accuracy of the CNN network declines to approximately 80\% at an ISNR of 8 dB, representing a 20\% reduction relative to clean samples.  (2) The proposed model demonstrates superior robustness despite the attack, achieving a recognition accuracy of approximately 95\% at 8 dB, thereby surpassing other models.   The proposed method utilises a differential attention mechanism to effectively differentiate components of the input attack, thereby diminishing their adversarial impact.


Fig. \ref{fig8} illustrates the impact of varying attack intensities on the proposed network's performance. The adversarial perturbations are systematically varied at four levels: 3/255, 6/255, 8/255, and 14/255. As the perturbation magnitude increases, the susceptibility of the network to adversarial samples becomes increasingly evident. Notably, when the perturbation reaches 14/255, the network's accuracy degrades significantly, dropping to approximately 40\% even under high ISNR  conditions. This observation underscores the substantial vulnerability of the network to adversarial attacks, highlighting that merely increasing the ISNR is insufficient to counteract the effects of such attacks.

Fig. \ref{fig9} (a) and \ref{fig9} (b) provide a comparative analysis of the prediction accuracies attained by the proposed consistent training and random masking training schemes across different attack coefficients.  Both training methodologies exhibit significant efficacy in improving the network's resilience to adversarial perturbations.  At an extreme attack perturbation level of 14/255 and an ISNR of 8 dB, the consistent training approach demonstrates a notable 25\% improvement in prediction performance, whereas the random masking training strategy results in a more pronounced 40\% enhancement.  The quantitative results validate the efficacy of both training paradigms and reveal the comparative advantages of random masking in reducing adversarial effects during high-intensity attack scenarios.  The observed performance improvements highlight the capacity of these training techniques to enhance network resilience against adversarial attacks, especially under difficult signal conditions.

\begin{figure}[htbp]
	\centering
	\subfigure[Differential transformer]{
		\includegraphics[width=3.5 in]{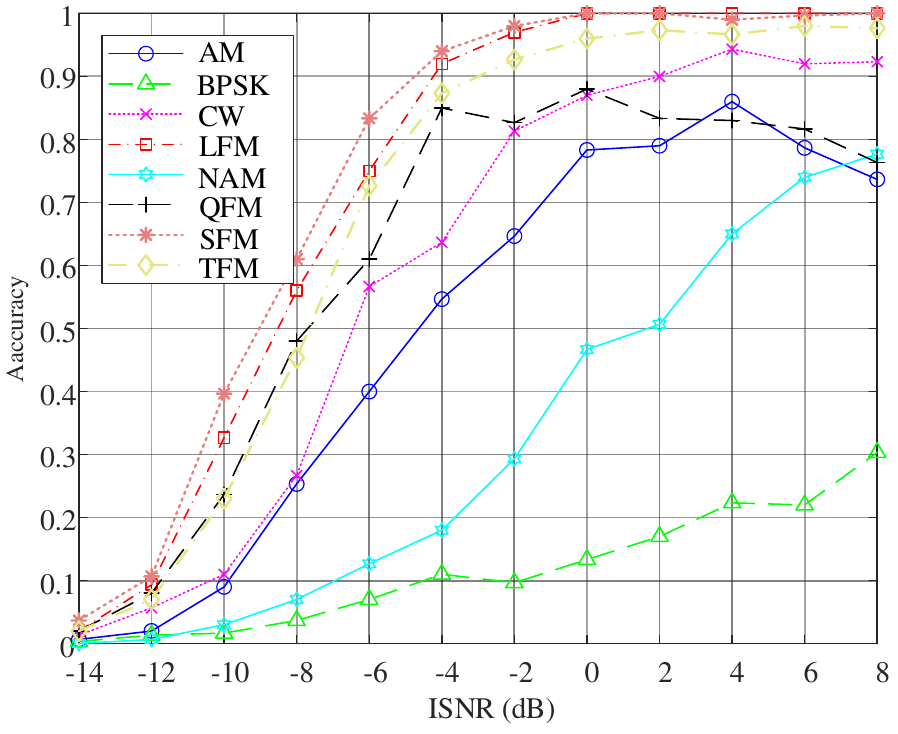}
		\label{fig10a}}
	\subfigure[Randomized masking training]{
		\includegraphics[width=3.5 in]{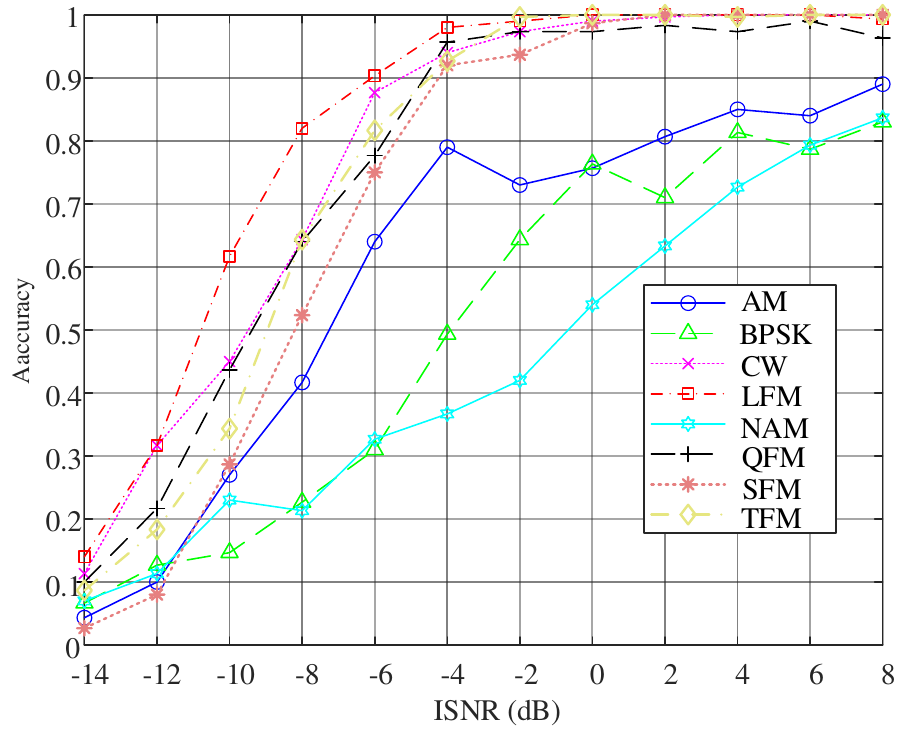}
		\label{fig10b}}%
	\caption{Recognition accuracy of each type of jamming with attack perturbation of 6/255. }
	\label{fig10}
\end{figure}

This study examines the prediction accuracy for each jamming type under two training conditions: with and without random occlusion, at an attack perturbation level of 6/255.  The comparative analysis, as shown in Fig. \ref{fig10}(a) and Fig. \ref{fig10}(b), indicates notable performance differences between the two training paradigms.  When trained without random occlusion, the network demonstrates significant vulnerabilities, as specific types of interference can effectively compromise the system, even in high ISNR conditions.  This leads to inadequate recognition performance, underscoring the shortcomings of conventional training methods in adversarial contexts.

The integration of random occlusion in training exhibits significant robustness.  Under uniform attack conditions, each type of jamming is effectively countered, with recognition rates consistently surpassing 80\% even at elevated ISNRs.  This improvement highlights the effectiveness of random occlusion as a defence mechanism, enhancing the network's resilience against various jamming attacks.  The performance disparity between the two training strategies highlights the importance of advanced training methods for ensuring reliable outcomes in adversarial contexts.

\begin{table}[]
	\caption{Comparison of different defense strategies.}
	\label{tab5}
	\setlength{\tabcolsep}{3pt}
	\begin{center}
	\begin{tabular}{|c|c|c|c|c|c|}
		\hline
		Method             & Clean  & 3/255  & 6/255  & 8/255  & 14/255 \\ \hline
		Defensive distillation\cite{b43}            & 0.8044 & 0.6682 & 0.5335 & 0.4470 & 0.2508 \\ \hline
		Feature denoising \cite{b44}           & 0.7965 & 0.6537 & 0.5173 & 0.4460 & 0.3007 \\ \hline
		BPFC\cite{b45}               & 0.7238 & 0.6095 & 0.5033 &   0.4332     &   0.3307     \\ \hline
		\begin{tabular}[c]{@{}c@{}}Proposed consistency \\ training\end{tabular}          & $\mathbf{0.8062}$ & 0.6907 & 0.5825 & 0.5255 & 0.4001 \\ \hline
		\begin{tabular}[c]{@{}c@{}}Proposed randomized \\ masking training\end{tabular} & 0.7987 & $\mathbf{0.7363}$ & $\mathbf{0.6647}$ & $\mathbf{0.6232}$ & $\mathbf{0.5239}$ \\ \hline
		\end{tabular}
\end{center}
\end{table}

Table \ref{tab5} presents a comparison of the proposed algorithm against various defence strategies, such as defensive distillation \cite{b43}, feature denoising \cite{b44}, BPFC \cite{b45}, the proposed consistent training, and the proposed random masking training.  To ensure a fair comparison, all defence strategies are implemented on the base model presented in Table \ref{tab2}.  The table indicates that both proposed defensive training strategies exhibit considerable advantages across all offensive perturbation factors. Both proposed methods achieve recognition accuracies of approximately 0.8 on clean samples, while demonstrating superior robustness across various perturbations compared to conventional defense approaches. Notably, the random masking training strategy exhibits significantly enhanced defensive capabilities. Extensive simulation results demonstrate the effectiveness of the proposed differential transformer in jamming signal identification, as well as the robustness of the two adversarial training strategies against adversarial samples.

\section{Conclusion}
In this paper, a differential transformer-based framework is presented, aimed at identifying wireless jamming in low-altitude wireless networks.  This study presents a new differential transformer-based network architecture that reduces the attention noise found in conventional transformer models, enhancing the network's ability to extract global features.  We propose a randomised masking training strategy that creates multiple feature extraction branches and implements various masking strategies for each branch, thereby minimising the influence of adversarial signals.  We present a consistent training strategy that improves adversarial robustness by utilising dual-branch regularisation to create robust and regular processing branches, while implementing a consistency loss to diminish the network's vulnerability to adversarial samples.  Simulation results indicate that the differential transformer network surpasses current methodologies, and both training strategies notably improve model robustness.

\end{document}